\documentclass[10pt,letterpaper]{article}
\usepackage[top=0.85in,left=2.75in,footskip=0.75in]{geometry}

\usepackage{amsmath,amssymb}

\usepackage{changepage}

\usepackage[utf8x]{inputenc}

\usepackage{textcomp,marvosym}

\usepackage{cite}

\usepackage{nameref,hyperref}

\usepackage[right]{lineno}

\usepackage{microtype}
\DisableLigatures[f]{encoding = *, family = * }

\usepackage[table]{xcolor}

\usepackage{array}

\newcolumntype{+}{!{\vrule width 2pt}}

\newlength\savedwidth

\raggedright
\usepackage[aboveskip=1pt,labelfont=bf,labelsep=period,justification=raggedright,singlelinecheck=off]{caption}

\makeatletter
\renewcommand{\@biblabel}[1]{\quad#1.}
\makeatother

\usepackage{lastpage,fancyhdr,graphicx}
\usepackage{epstopdf}
\fancyheadoffset[L]{2.25in}
\fancyfootoffset[L]{2.25in}
\begin{document}
\vspace*{0.2in}

\begin{flushleft}
{\Large
\textbf\newline{DMRIntTk: integrating different DMR sets based on density peak clustering} 
}
\newline
\\
Wenjin Zhang\textsuperscript{1},
Wenlong Jie\textsuperscript{2},
Wanxin Cui\textsuperscript{2},
Guihua Duan\textsuperscript{2},
You zou\textsuperscript{2,3*},
Xiaoqing Peng\textsuperscript{1*}
\\
\bigskip
\textbf{1} Center for Medical Genetics \& Hunan Key Laboratory of Medical Genetics, School of Life Sciences, Central South University, 410083, Changsha, China
\\
\textbf{2} Hunan Key Laboratory of Bioinformatics, School of Computer Science and Engineering, Central South University, 410083, Changsha, China
\\
\textbf{3} High Performance Computing  Center, Central South University, 410083, Changsha, China
\bigskip

%
%





* zouyou@csu.edu.cn; xqpeng@csu.edu.cn

\end{flushleft}
\section*{Abstract}
\textbf{Background}: Identifying differentially methylated regions (DMRs) is a basic task in DNA methylation analysis. However, due to the different strategies adopted, different DMR sets will be predicted on the same dataset, which poses a challenge in selecting a reliable and comprehensive DMR set for downstream analysis.

\noindent \textbf{Results}: Here, we develop DMRIntTk, a toolkit for integrating DMR sets predicted by different methods on a same dataset. In DMRIntTk, the genome is segmented into bins and the reliability of each DMR set at different methylation thresholds is evaluated. Then, the bins are weighted based on the covered DMR sets and integrated into DMRs by using a density peak clustering algorithm. To demonstrate the practicality of DMRIntTk, DMRIntTk was applied to different scenarios, including different tissues with relatively large methylation differences, cancer tissues versus normal tissues with medium methylation differences, and disease tissues versus normal tissues with subtle methylation differences. The results show that DMRIntTk can effectively trim the regions with small methylation differences in the original DMR sets and therefore it  can enhance the proportion of DMRs with higher methylation differences. In addition, the overlap analysis suggests that the integrated DMR sets are quite comprehensive, and the functional analysis indicates the integrated disease-related  DMR sets are significantly enriched in biological pathways, which are associated with the pathological mechanisms of the diseases.

\noindent \textbf{Conclusions}: Conclusively, DMRIntTk can help researchers obtaining a reliable and comprehensive DMR set from many prediction methods.

\noindent \textbf{Keywords}:{Differentially methylated regions, Methylation array, Cancer-related differentially methylated regions, Tissue-specific differentially methylated regions, Density peak clustering.}



\newpage

\section{Background}\label{sec1}
DNA methylation is a crucial epigenetic modification that plays a pivotal role in various biological processes, such as tissue differentiation, embryonic development, tumorigenesis and aging. Identifying differentially methylated regions (DMRs) is a fundamental task in DNA methylation analysis. The identified tissue-specific\cite{Gai2018,Groleau2021}, cell type-specific\cite{Wiencke2016,Adams2023} and disease-associated DMRs\cite{Watson2016,Peipei2019,Thanh2023} can help to investigate the underlying molecular mechanisms of differentiation and pathogenesis, and serve as potential methylation biomarkers for screening diseases.

Many methods have been developed for detecting DMRs on methylation array data. These methods can be broadly categorized into CpG-based and candidate-region-based methods. In CpG-site-based methods\cite{Xu2021,Timothy2021,Tamar2013,Linghao2017,Brent2012,Lee2015}, differentially methylated CpGs (DMCs) are first identified based on the methylation levels of the samples between two groups. The p-values of DMCs are often corrected based on the auto-correlations of CpG site, the correlations between adjacent CpG sites, the p values of a adjacent CpG sites, etc. Then adjacent DMCs are agglomerated to form DMRs if some defined criteria are satisfied, such as a minimum distance between neighboring DMCs.

In candidate-region-based methods\cite{Martin2014,Fabian2019,Charles2013,Lissette2019,Dan2012,Jordi2019,Andrew2012,Yan2011,Raivo2016}, there are typically two types of candidate regions: sample-independent regions and sample-dependent regions. The sample-independent regions are either predefined based on functional regions, such as CpG islands and shores, or generated by a sliding window on the genome. The sample-dependent regions are generated according to the characteristics of samples, including the coverage, depth of CpG sites, and methylation levels of CpG sites, or the methylation changes of CpG sites among multiple samples. Once the candidate regions are determined, DMRs are then identified by comparing the methylation levels of these regions across different samples.

Due to the different strategies used in different methods, different DMR sets are predicted on a same dataset, in terms of the DMR length, the number of probes and CpG sites included, and the methylation differences. There is no method that can perform well on datasets in all scenarios, and we can hardly figure out which scenario a method is most applicable to. Therefore, it is challenging to select a desirable DMR set for downstream analysis.

As we know, different methods have their own advantages in predicting different types of DMRs, and DMRs that are detected by most methods and with relatively large methylation differences can be regarded as highly reliable DMRs. Therefore, we develop a toolkit, DMRIntTk, which evaluates the reliability of different DMR sets and integrates them using a density peak clustering (DPC) algorithm. To evaluate the performance of DMRIntTk, it was applied to the DMR sets in four representative scenarios, including datasets with large methylation differences (five different tissues), medium methylation differences (the prostatic cancer (PCa) tissues versus the adjacent normal prostate tissues, and the benign versus other five histological stages of PCa tissues), and small methylation differences (brain regions of patients with Alzheimer's disease (AD) versus the normal ones). The results show that DMRIntTk can enhance the proportion of DMRs with higher methylation differences. In addition, the overlap analysis suggests that the integrated DMR set is more comprehensive and reliable than individual DMR sets.

\section{Methods}\label{sec2}
\subsection{The pipeline of DMRIntTk}\label{subsec21}
\begin{figure*}
    \centering
    \includegraphics[scale=0.30]{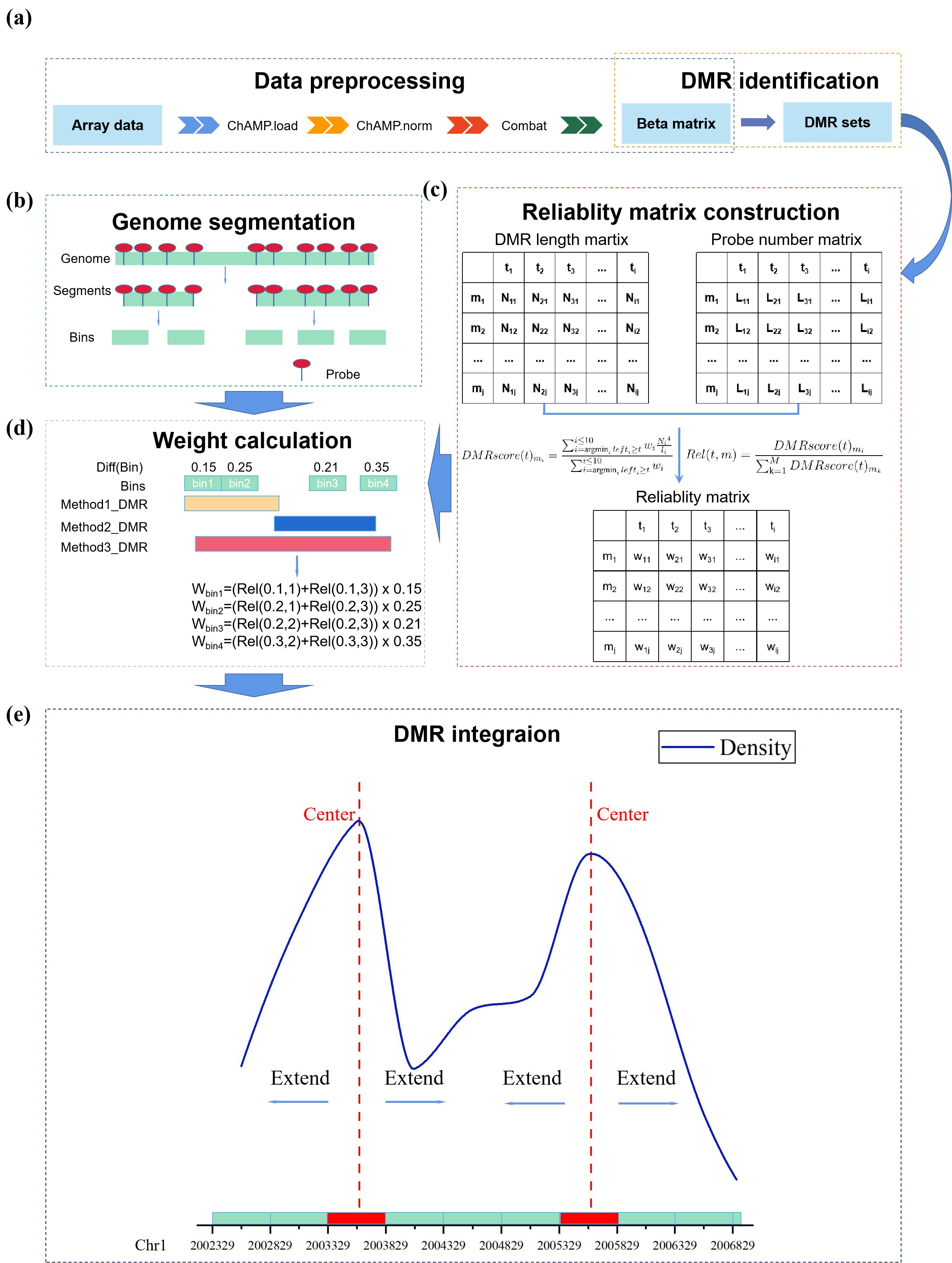}
    \caption{A schematic diagram of DMRIntTk. (a) Data pre-processing and DMR identication steps output DMR sets are used as standard input in DMRIntTk. There are four major parts in DMRIntTk, including (b) segmenting the genome, (c) constructing the reliability matrix, (d) weighting bins and (e) integrating bins based on density peak clustering. The example of each part are shown.}\label{fig1}
\end{figure*}

In this paper, a toolkit, DMRIntTk, is developed to integrate the DMR sets predicted by different methods. The pipeline of DMRIntTk  mainly contains four steps, including segmenting the genome, constructing the reliability matrix, weighting the bins and integrating DMRs, as shown in Figure \ref{fig1}. The details of each step is described in following subsections.

\subsection{Construct reliability matrix}\label{subsec22}
Since different methods have their own preferences and advantages in predicting DMRs with different properties, the reliability of DMR sets predicted by different methods should be fully accessed. Therefore, we propose a metric $DMRscore$ to evaluate the reliability of different DMR sets.

Let $m_{max}$ denote the maximum methylation difference of DMRs. Then, divide the range of methylation difference [0, $m_{max}$] into ten intervals with equal length $m_{max}/10$. For each DMR set predicted by a method, the total number of probes and the total length of DMRs with methylation differences falling in each interval are calculated, respectively. Given a certain value of a threshold $t$ in the range of [0, $m_{max}$], the $DMRscore$ of a DMR set $m_k$ is calculated as Equation \ref{eq1}.

\begin{equation}
DMR score(t)_{m_k}=\frac{{\sum\limits_{i=\mathop{\mathrm{argmin}}\limits_{i}{left_{i}\geq{t}}}^{i\leq{10}}}w_{i}\frac{{N_i}^4}{l_i}}{\sum\limits_{i=\mathop{\mathrm{argmin}}\limits_{i}{left_{i}\geq{t}}}^{i\leq{10}}w_{i}}\label{eq1}
\end{equation}

Where $i$ denotes the $i$-th interval, $left_i$ and $right_i$ denote the left and right points of the $i$-th interval, $w_i$ denotes the weight of the $i$-th interval, set as $\frac{left_i+right_i}{2}$, $N_i$ and $l_i$ denote the total number of probes and the total length of DMRs with methylation differences falling in the $i$-th interval, respectively.

With a certain threshold $t$, if the $DMRscore$ of a DMR set is greater than that of other DMR sets, it indicates that it is more reliable and comprehensive with a higher probe density and more probes in DMRs whose methylation differences are greater than $t$. With $M$ DMR sets and a given threshold $t$, the normalized reliability score of each DMR set $m_k$, denoted as $Rel(t,m_k)$, can be calculated according to Equation \ref{eq2}.
\begin{equation}
Rel(t,m_k)=\frac{DMRscore(t)_{m_k}}{\sum\limits_{n=1}^{|M|}{DMRscore(t)_{m_n}}}\label{eq2}
\end{equation}

The reliability scores of $M$ DMR sets can be calculated for each value of $t$ in the set \{0, $m_{max}/10$, $2*m_{max}/10$, ..., $9*m_{max}/10$\}. Then, a reliability matrix $R$ with $M \times 10$ can be constructed.

\subsection{Segment the genome and weight the bins}\label{subsec23}
For methylation array data, the whole genome is segmented into non-overlapping genomic bins based on the distances of probes. Firstly, segments are formed by agglomerating the adjacent probes with a genomic distance less than 500bp. Then, a window size of 500 bp is adopted to generate bins from the segments. For 450K array, 10,9524 bins are generated, and for 850K array, 17,9783 genomic bins are generated.

To obtain a reliable and comprehensive DMR set from the predictions of multi methods, each bin is evaluated by considering the methylation difference and the reliability scores of the covering DMR sets. For a bin $B_i$, let $Dif(B_i)$ denotes the absolute value of the methylation difference between two groups on $B_i$, and $N$ denotes the number of DMR sets covering $B_i$. Then, the weight of $B_i$ can be calculated by integrating the reliability scores of $N$ DMR sets and the methylation difference $Dif(B_i)$, as defined in Equation \ref{eq3}.
\begin{equation}
w(B_i)={\sum\limits_{k=1}^{|N|}}Rel(Dif(B_i),m_k)\times{Dif(B_i)}\label{eq3}
\end{equation}

\subsection{Generate integrated DMRs based on density peak clustering}\label{subsec24}
Bins covered by more DMR sets and with higher methylation differences will be assigned with higher weights, which are the basic elements in generating integrated DMRs. We applied an adapted density peak clustering (DPC) algorithm on weighted bins to identify DMRs. In this algorithm, the bins are treated as the basic points. The local density is defined as the weight of a bin, and the distance between two bin is defined as their genomic distance.

Assume that there two thresholds, $c_t$  and $n_t$, which are used for identifying cluster centers and cluster members, respectively. Firstly, bins with weights greater than $c_t$ are identified as cluster centers. Then, sequential adjacent cluster centers will be merged into one cluster center. For each bin which is not a cluster center and between two cluster centers, it will assigned to the cluster with a smaller genomic distance. Then pre-defined regions are formed by cluster centers and the assigned bins.

For each pre-defined region, a cluster will be identified by a cluster center extension mode. A cluster is initialized as the cluster center in a pre-defined region. Then, for the nearest bin in the pre-defined region from each side of the cluster, if its weight is greater than $n_t$, it will be merged into the cluster and the extension on this side will be continued until the weight of the next nearest bin is not greater than $n_t$. The finally formed cluster is considered as an integrated DMR and its methylation difference is calculated as the mean  methylation difference of all clustered bins.

\subsection{The DMR set integrated by DMRIntTk}\label{subsec25}
Since there are two parameters in the DPC algorithm to identify DMRs, different settings of $c_t$ and $n_t$ will result in different integrated DMR sets. To facilitate the researchers and make the usage simple, DMRIntTk provides an automatic mechanism for determining the parameter values and outputting the final integrated DMR set.

Given that $m_{max}$ denote the maximum methylation difference of DMRs, the value of $c_t$ will be enumerated from the set of \{0.2*$m_{max}$,...,1*$m_{max}$\} and the value of $n_t$ will be enumerated from the set of \{0.1*$m_{max}$,...,0.9*$m_{max}$\}. For each combination of  $c_t$ and $n_t$, the DPC algorithm is applied to obtain an integrated DMR set. Then, for each integrated DMR set, the total number of contained probes, the total length of DMRs, and a proportion of DMRs with methylation differences greater than $0.5*m_{max}$ are calculated. Based on these information, the final  integrated DMR set output by DMRIntTk will satisfy the following two conditions: (1) contains a higher proportion of DMRs with methylation differences greater than $0.5*m_{max}$ than any individual DMR set; (2) has the most probes or the longest total DMR length than other integrated DMR sets.

\section{Results}\label{sec3}
In this study, seven state-of-the-art DMR detection methods, including Bumphunter\cite{Jaffe2012}, ProbeLasso\cite{Butcher2015}, DMRcate\cite{Peters2015}, comb-p\cite{Pedersen2012}, ipDMR\cite{Xu2020}, mCSEA\cite{Martorell2019}, and seqlm\cite{Raivo2016}, were applied to predict DMRs on DNA methylation 450K array data in four scenarios.  For the parameters in each method, the default values were adopted. For each DMR set,  DMRs with adjusted p-values greater than 0.05 or containing less than 3 probes were filtered out. Finally, DMRIntTk was applied to integrate the original DMR sets predicted by seven methods.

To evaluate the integrated DMR sets obtained by DMRIntTk, the methylation difference distribution of the integrated DMR sets and  the original DMR sets were compared. Further, an overlap analysis was carried between the integrated DMR sets and the original DMR sets. Moreover, the enrichment analysis of functional pathways was analyzed.

\subsection{Materials}\label{subsec31}
In this paper, methylation array datasets of four scenarios are involved, including 1) five tissues, 2) the PCa tissues versus the adjacent normal prostate tissues, 3) the benign versus other histological stages of PCa tissues, and 4) the brain tissues with AD versus the normal brain tissues. The DMRIntTk is used to integrate DMR sets predicted by different method on the methylation array datasets in these scenarios.

In scenario 1, five tissues were involved, including 5 liver tissues, 6 muscle tissue and 6 omentum tissues extracted from GEO with accession id GSE48472\cite{Roderick2013}, and 5 lymphoid tissues and 7 tonsil tissues from GSE50192\cite{Kaie2014,Natalia2016}. In scenario 2, the 450K methylation profiles of 31 PCa tissues and 16 adjacent normal prostate tissues were extracted from GSE112047\cite{Erfan2018}. In scenario 3, the 450K methylation profiles of PCa tissues under different histological stages were extracted from GSE157272\cite{Romina2021}, which contains 10 benign prostate tissues, 7 proliferative inflflammatory atrophy (PIA) prostate tissues, 6 high grade prostatic intra-epithelial neoplasia (HGPIN) tissues, 7 Indolent PCa tissues, 8 aggressive PCa tissues and 6 metastatic PCa tissues. In scenario 4, the 450K methylation profiles of four brain regions of patients with Alzheimer's disease (AD) and normal controls were involved, including the entorhinal cortex (EC) (with 146 AD samples and 97 normal samples extracted from GSE43414\cite{Pidsley2013}, GSE105109\cite{Smith2019} and GSE125895\cite{Semick2019}), the frontal cortex (FC) (with 171 AD and 118 normal samples extracted from GSE43414, GSE66351\cite{Gasparoni2018}, and GSE80970\cite{Smith2018}, and 308 AD and 233 normal samples extracted from the Religious Orders Study and Memory and Aging Project (ROSMAP) cohort\cite{Bennett2018} with Synapase ID syn3219045), and the superior temporal gyrus (STG) (with 135 AD samples and 96 normal samples extracted from GSE43414 and GSE80970), and the hippocampus (HP) (with 17 AD samples and 48 normal samples extracted from GSE125895).

All the 450K datasets were preprocessed with quality control and normalized by using the ChAMP R package\cite{Tiffany2014}. Batch effects were removed by using the Combat function in the sva R package\cite{Jeffrey2012}, and covariates including sex and age were adjusted using the linear mixed model.

\subsection{Performance of the DMRIntTk on different tissue pairs}\label{subsec32}
\begin{figure*}
    \centering
    \includegraphics[scale=0.21]{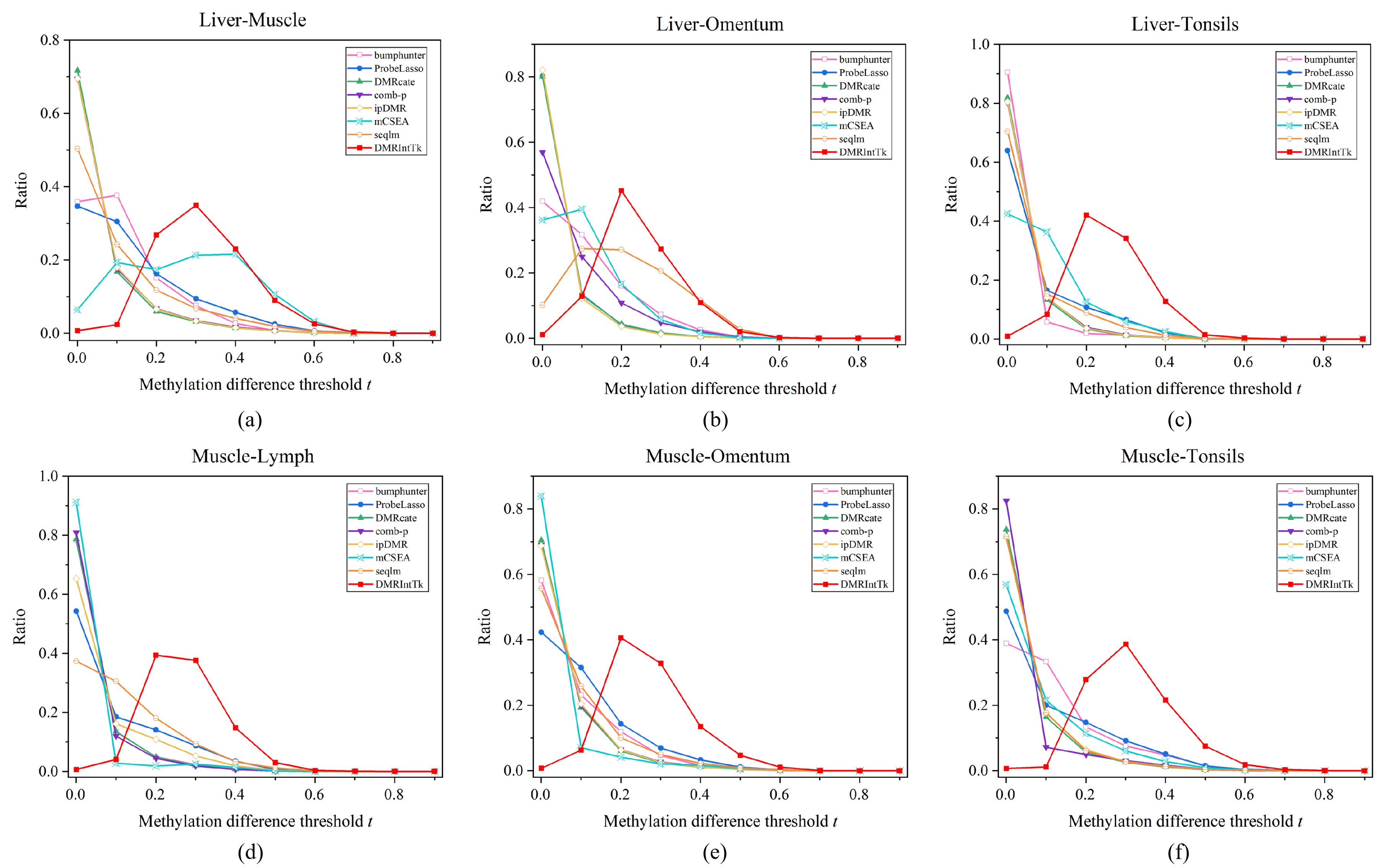}
    \caption{The methylation difference distributions of DMR sets on six pairs of tissues. (a) The liver vs. the muscle. (b) The liver vs. the omentum. (c) The liver vs. the tonsils. (d) The muscle vs. the lymph. (e) The muscle vs. the omentum. (f) The muscle vs. the tonsils.}
    \label{fig2}
\end{figure*}

Seven methods were applied to different scenarios to identify DMRs, and the DMR sets predicted on a same dataset were integrated by DMRIntTk. The methylation difference distributions of seven original DMR sets and the integrated one on each pair of tissues are illustrated, as shown in Figure \ref{fig2}. The range of methylation difference [0,1] is divided into intervals with equal length, as shown in X axis.  Y axis denotes the ratio of DMRs with methylation differences falling in each interval to the total number of DMRs in each DMR set. It can be found that the methylation differences of majority of DMRs between a pair of tissues are less than 0.6, and there are large proportions of DMRs with methylation differences less than 0.2. Compared with these original DMR sets, it can be observed that when the methylation difference threshold $t$ is in the range of [0.2, 0.5], the ratios of DMRs in the integrated set are significantly greater than those in the original DMR sets.
It suggests that DMRIntTk effectively trim the regions with small methylation differences in original DMR sets by segmenting the genome into bins and weighting the bins, therefore enhances the proportion of DMRs with medium methylation differences.

To demonstrate that an integrated DMR set is more comprehensive than a single DMR set and contains high reliable DMRs, an overlap analysis was carried between the integrated DMR sets and the original DMR sets. To reduce the bias introduced by DMRs with small methylation differences,  only the DMRs with methylation differences not less than $0.5*m_{max}$ are involved in analysis, where $m_{max}$ denotes the maximum methylation difference in DMRs between two groups. With the overlap length calculated between an integrated DMR set and an original DMR set, two overlap rates are calculated against the total lengths of DMRs (with methylation differences $\geq 0.5*m_{max}$) in the integrated DMR set and the original DMR set, respectively.
\begin{figure*}
    \centering
    \includegraphics[scale=0.21]{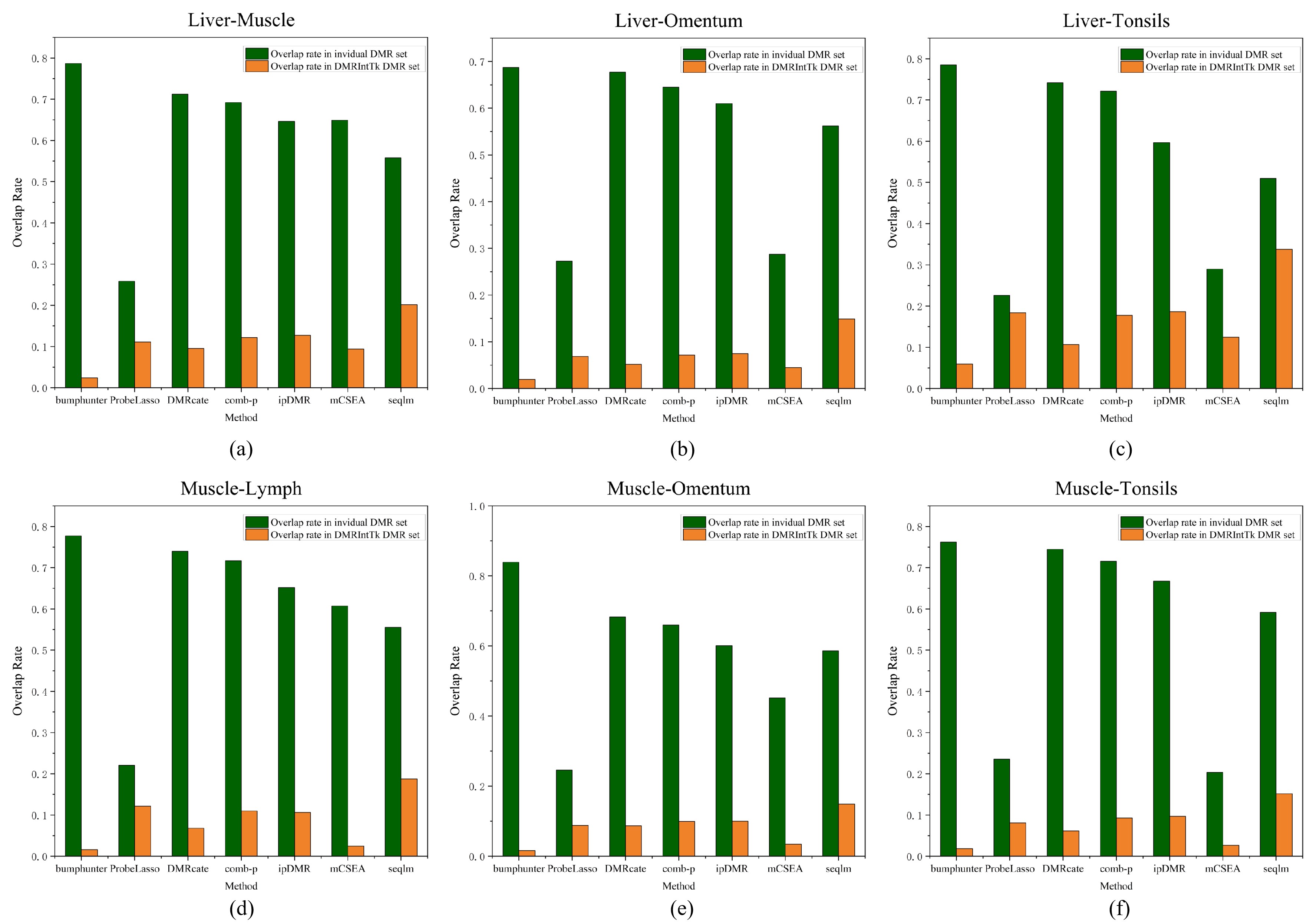}
    \caption{The overlap rates between the integrated DMR set and individual DMR sets on six pairs of tissues. (a) The liver vs. the muscle. (b) The liver vs. the omentum. (c) The liver vs. the tonsils. (d) The muscle vs. the lymph. (e) The muscle vs. the omentum. (f) The muscle vs. the tonsils. Overlap rate in individual DMR set denotes the overlap rate calculated against the total lengths of DMRs (with methylation differences $\geq 0.5*m_{max}$) in an original DMR set, and Overlap rate in DMRInTK DMR set denotes the overlap rate calculated against the total lengths of DMRs (with methylation differences $\geq 0.5*m_{max}$) in the integrated DMR set.}
     \label{fig5}
\end{figure*}

As shown in Figure \ref{fig5}, for each pair of tissues, it can be observed that majority of overlap rates calculated based on the lengths of original DMR sets are greater than 0.6, which indicates that most DMRs with methylation differences $\geq 0.5*m_{max}$ in these original DMR sets are retained in the integrated DMR set. Further, it can be found out that the overlap rates calculated based on the length of the integrated DMR set are mainly less than 0.2, which indicates the regions with high methylation differences from different DMR sets are effectively integrated and the DMR sets integrated by DMRIntTk are quite comprehensive.

\subsection{Performance of the DMRIntTk on PCa versus adjacent prostate tissues}\label{subsec33}

\begin{figure*}
    \includegraphics[scale=0.21]{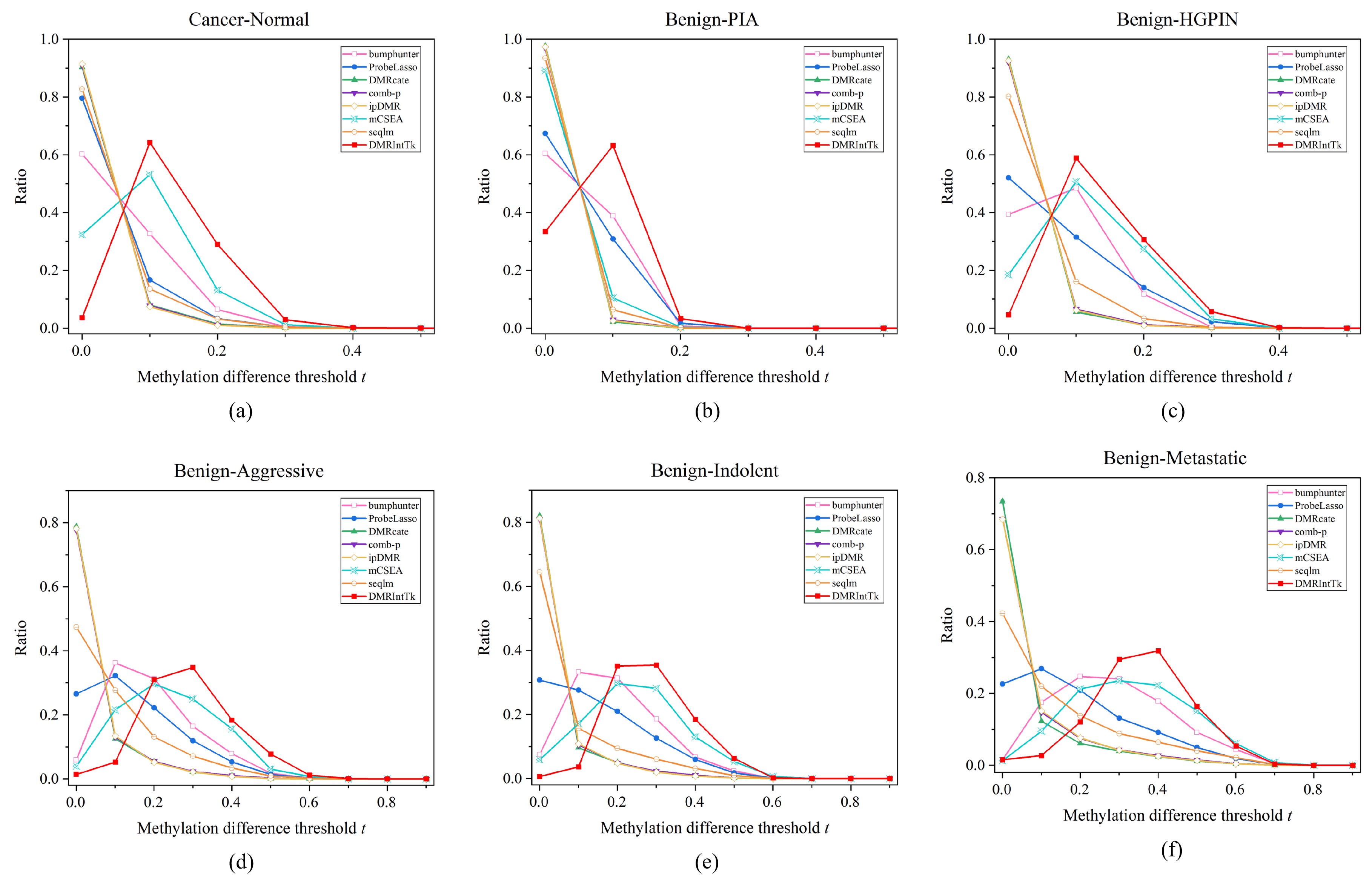}
    \caption{The methylation difference distributions of DMR sets predicted between PCa and adjacent normal prostate tissues, and between the benign and five other histological stages of PCa. (a) PCa v.s. Normal. (b) Benign v.s. PIA. (c)Benign v.s. HGPIN. (d) Benign v.s. Indolent. (e) Benign v.s. Aggressive. (f) Benign v.s. Metastatic.}
    \label{fig3}
\end{figure*}

The methylation difference distributions of DMR sets between the PCa tissues and adjacent normal tissues are compared, as shown in Figure \ref{fig3}(a).  It can be observed that the methylation differences of majority of DMRs between PCa and adjacent prostate tissues are less than 0.3. It can be figured out that when methylation difference $t$ is in the range of [0.1, 0.3], the ratios of DMRs in the integrated DMR set are significantly greater than those in the original DMR sets.

\begin{figure*}
    \centering
    \includegraphics[scale=0.21]{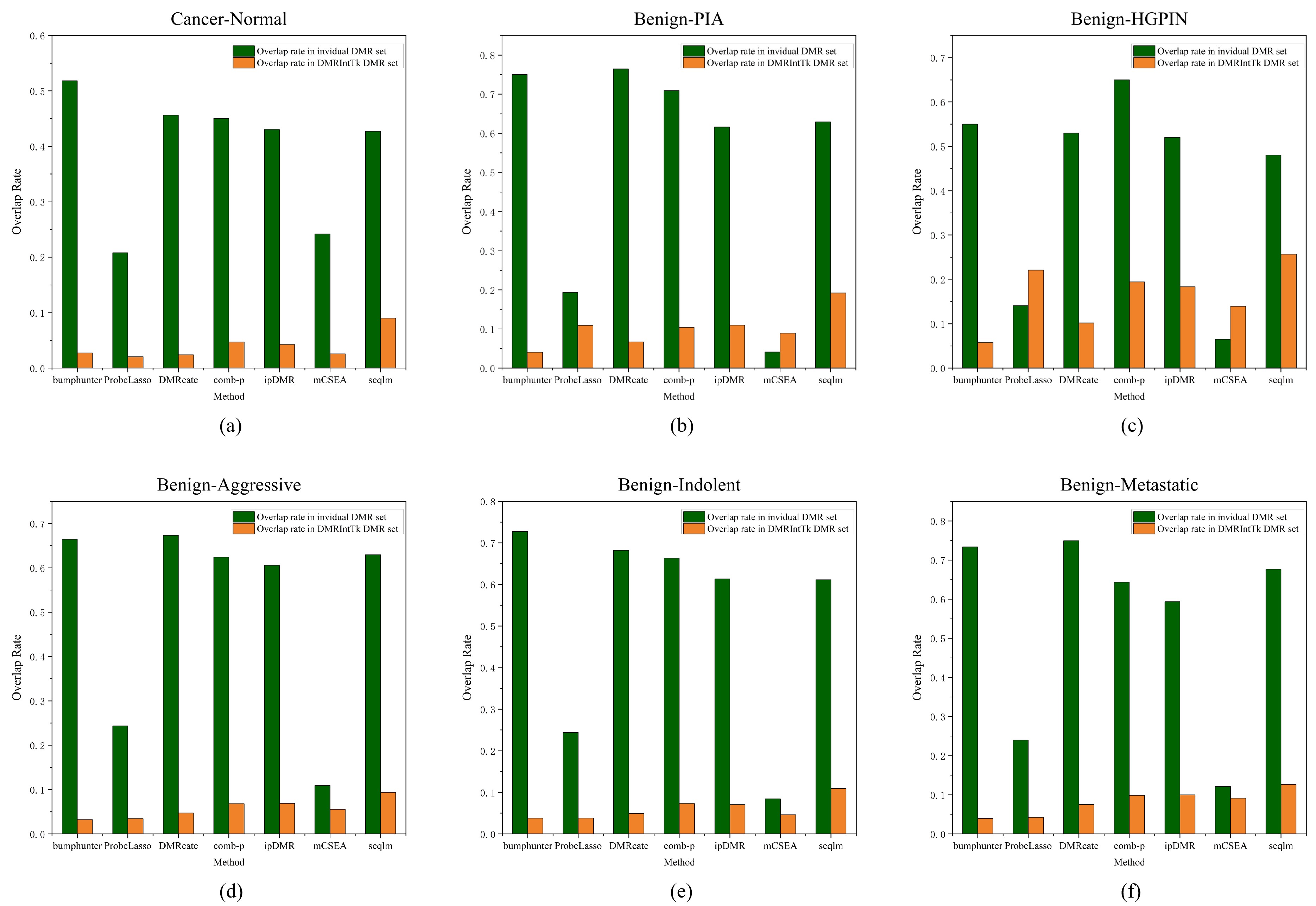}
    \caption{The overlap rates between the integrated DMR set and individual DMR sets predicted between PCa and adjacent normal prostate tissues, and between the benign and five other histological stages of PCa.
    (a) PCa v.s. Normal. (b) Benign v.s. PIA. (c) Benign v.s. HGPIN. (d) Benign v.s. Indolent. (e) Benign v.s. Aggressive. (f) Benign v.s. Metastatic. Overlap rate in individual DMR set denotes the overlap rate calculated against the total lengths of DMRs (with methylation differences $\geq 0.5*m_{max}$) in an original DMR set, and Overlap rate in DMRInTK DMR set denotes the overlap rate calculated against the total lengths of DMRs (with methylation differences $\geq 0.5*m_{max}$) in the integrated DMR set.}
     \label{fig6}
\end{figure*}

The overlap rates between the integrated DMR set and the original DMR sets on the PCa and normal adjacent prostate tissues  are compared, as shown in Figure \ref{fig6}(a). It can be found out that the overlap rates calculated against the length of original DMR sets are above 0.4, except for ProbeLasso and mCSEA, while the overlap rates calculated against the integrated DMR set are less than 0.1.

\subsection{Performance of the DMRIntTk on benign versus other histological stages of PCa tissues}\label{subsec34}

The methylation difference distributions of DMR sets between the benign and five other histological stages of PCa are illustrated, as shown in Figure \ref{fig3}(b)-(f). It can be found out that, the methylation differences of DMRs identified by senven methods between the Benign and the PIA are less than 0.2, while DMRIntTk enhances the proportion of DMRs with methylation differences in the range of [0.1, 0.2). On the Benign v.s. the HGPIN, the ratio of DMRs integrated by DMRIntTk is higher than those in the original DMR sets when $t\geq0.1$. On the Benign v.s. the Aggressive, the Benign v.s. the Indolent, and the Benign v.s. the Metastatic, the ratios of DMRs with methylation differences $\leq 0.1$ predicted by DMRcate, comb-p, seqlm, and ipDMR, are up to 0.8, while bumphunter, mCSEA, and Probelasso predict higher ratios of DMRs with methylation differences in the range of [0.1, 0.4]. On these four pairs, the methylation differences of integrated DMRs are mainly distributed in the range of [0.2, 0.5], and the corresponding ratios are greater than those in the original DMR sets.

For the benign and five other histological stages of PCa, as shown in Figure \ref{fig6}(b)-(f), we can observe that the overlap rates of most original DMR sets, except for ProbeLasso and mCSEA, are above 0.5  on the benign v.s. the HGPIN, and  above 0.6 on the benign v.s. the PIA, the benign v.s. the aggressive, the benign v.s. the Indolent, and the benign v.s. metastatic. It suggests that nearly half of the DMRs with methylation differences $\geq 0.5*m_{max}$ in these original DMR sets are included in the integrated DMR set. The overlap rates calculated against the integrated DMR sets are less than 0.2 on the benign v.s. the PIA and  the benign v.s. the HGPIN, and less than 0.1 on  the benign v.s. the aggressive, the benign v.s. the Indolent, and the benign v.s. metastatic. It further validates that the integrated DMR sets are quite comprehensive and very different from the original DMR sets.

It can be observed that  the overlap rates between the DMR sets predicted by mCSEA and the integrated DMR sets are low and the ones calculated against the mCSEA DMR set on Benign v.s. PIA and Benign v.s. HGPIN are  higher than those calculated against the integrated DMR sets. The reason is that mCSEA identifies DMRs based on the predefined functional regions, some of which are covered by sparse probes and the bins in these regions are not covered by any probes and the corresponding weights are zeros. Therefore, these bins are not integrated by DMRIntTk.

\subsection{Performance of the DMRIntTk on the AD versus normal brain tissues}\label{subsec35}

\begin{figure*}
    \centering
    \includegraphics[scale=0.30]{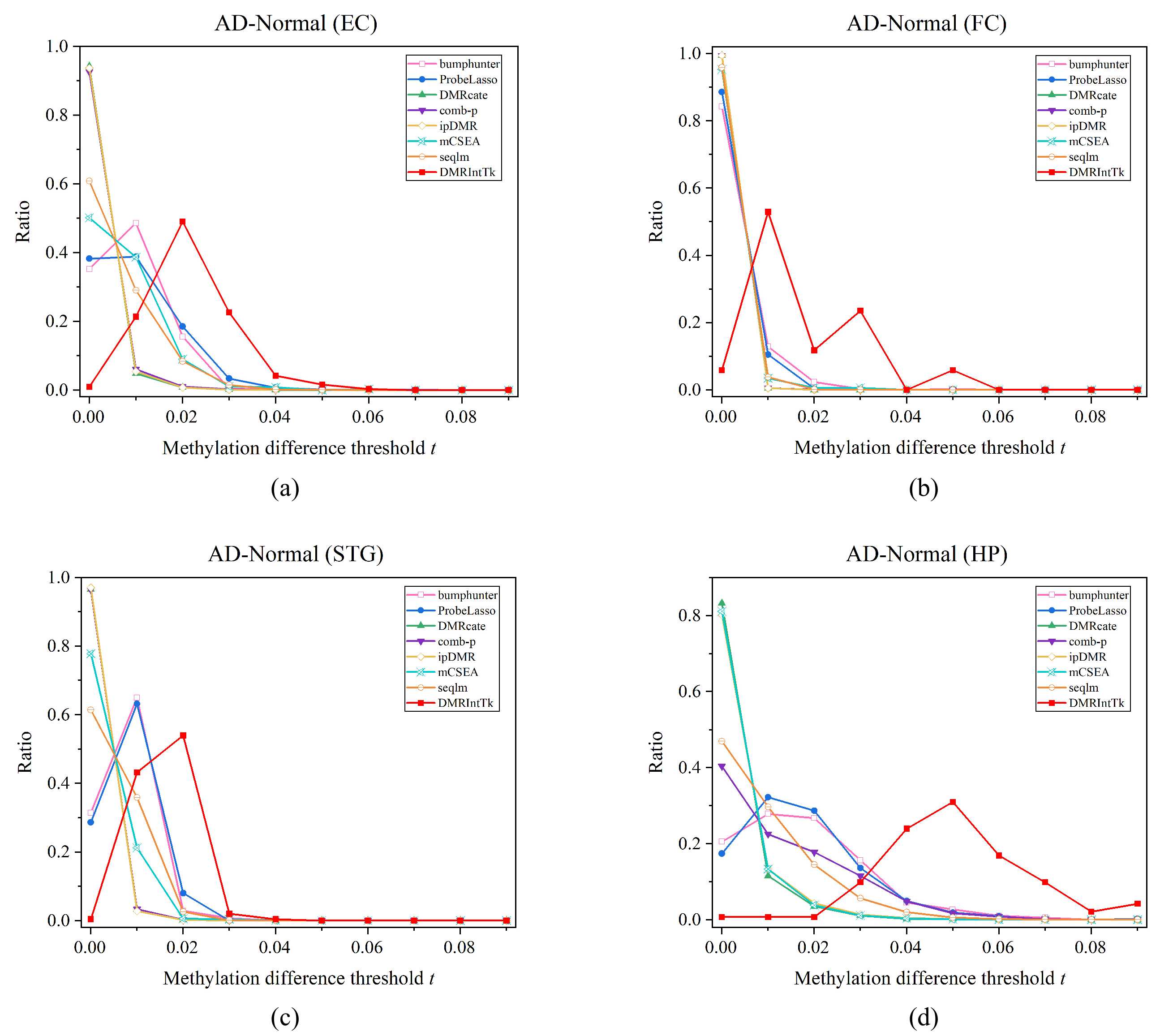}
    \caption{The distribution of methylation differences of DMR sets between the AD patients and normal controls on four brain regions. (a) entorhinal cortex (EC), (b) frontal cortex (FC), (c) hippocampus (HP) and (d) superior temporal gyrus (STG).}
    \label{fig4}
\end{figure*}

As shown in Figure \ref{fig4}, it can be found that the methylaiton differences between AD patients and normal controls on four brain regions are quite small, less than 0.1. The ratio peaks of most original DMR set are mainly in the range of (0, 0.01], while that of the integrated DMR sets are in the range of [0.02, 0.05]. It can be found out that DMRIntTk gains higher ratios of DMRs with methylation differences in the range of [0.02, 0.04], [0.01, 0.05], [0.02, 0.03] and [0.04, 0.09]  than other methods between AD patients and normal controls on the EC, FC, STG and HP, respectively.

\begin{figure*}
    \centering
    \includegraphics[scale=0.30]{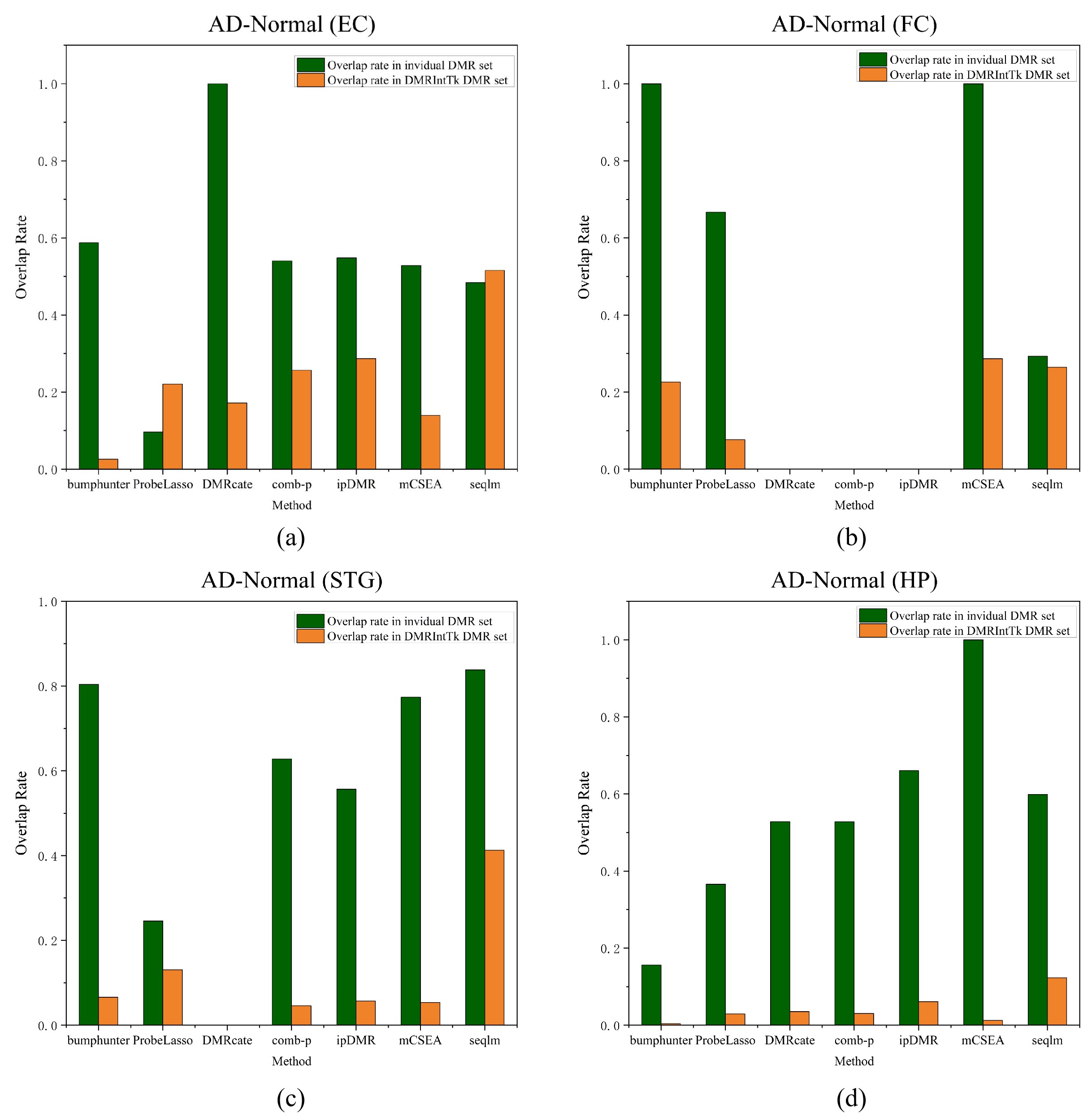}
    \caption{The overlap rates between the integrated DMR set and individual DMR sets on the four brain regions of the AD patients and normal controls. (a) entorhinal cortex (EC). (b) frontal cortex (FC). (c) hippocampus (HP). (d) superior temporal gyrus (STG).}
     \label{fig7}
\end{figure*}

For the DMRs predicted from AD patients and normal controls on four brain regions, the overlap analysis between the original DMR sets and the integrated ones is illustrated, as shown in Figure \ref{fig7}. It can be found out that the overlap rates calculated against the original DMR sets are significantly higher than those calculated against the integrated DMR sets, except for the ProbeLasso and seqlm in the EC. It is noteworthy that all DMRs with methylation differences $\geq 0.5*m_{max}$ in the DMR set predicted by DMRcate in the EC, by bumphunter in the FC, and by mCSEA in the FC and the HP are effectively integrated by  DMRIntTK, in which the overlap rates against the original DMR sets are near 1. Since comb-p and DMRcate predict only one DMR with methylation differences $\geq 0.5*m_{max}$ in the FC and the STG, repectively, the corresponding overlap rates between the DMR sets predicted by comb-p and DMRcate are zero. DMRcate and ipDMR predict none DMRs on the FC, and therefore there are no corresponding overlap rates.

\subsection{Functional Pathway Analysis of the Integrated DMR sets}\label{subsec36}
To analyze the function of the integrated DMR sets, firstly, the DMR located genes (DMGs) are extracted. Then, the GO enrichment analysis is performed on these DMGs by David\cite{dennis2003david}.

\subsubsection{GO enrichment analysis of the integrated DMR set between PCa tissues and adjacent normal tissues}\label{subsec361}
\begin{figure*}
    \centering
    \includegraphics[scale=0.45]{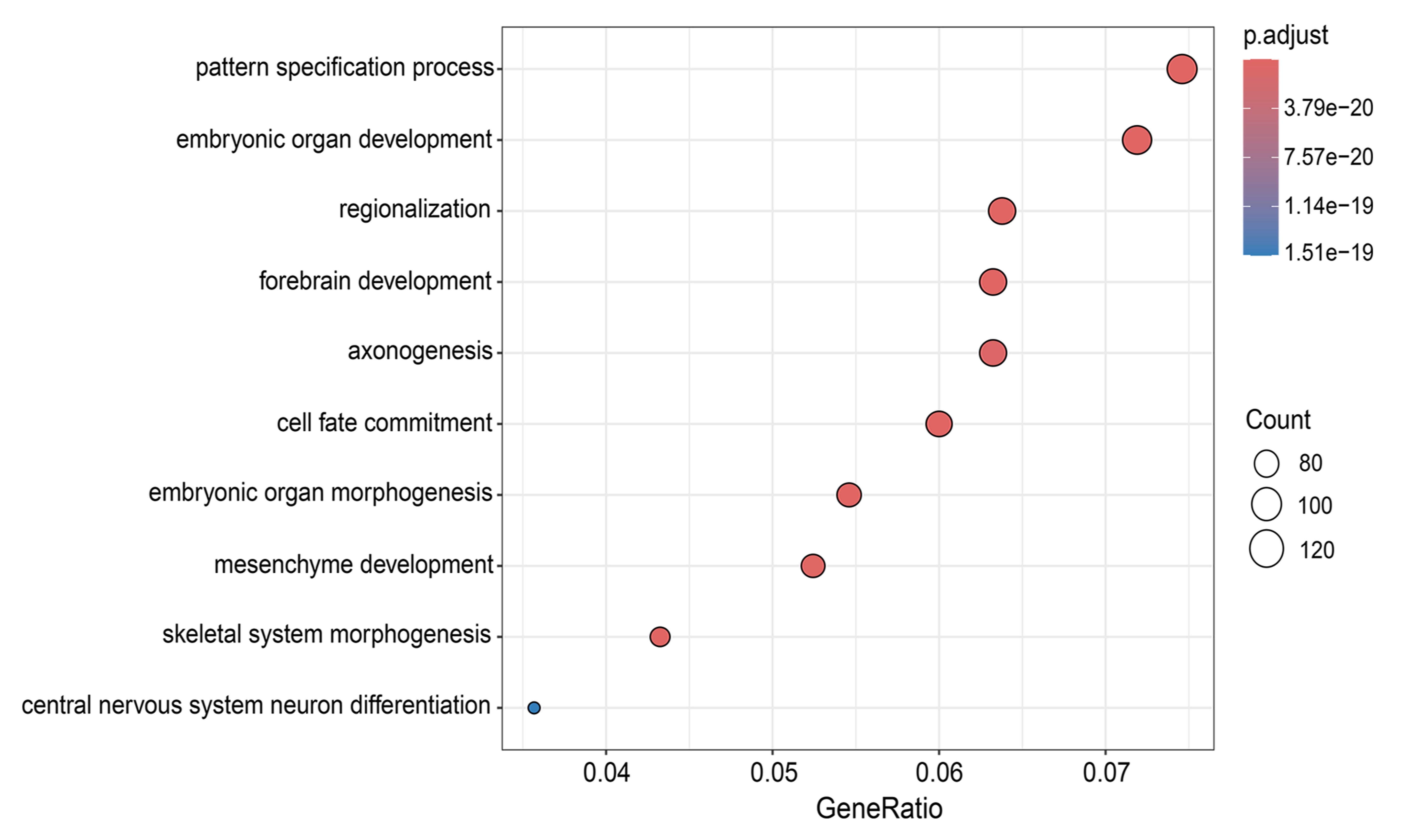}
    \caption{GO enrichment analysis of the integrated DMR set on PCa versus normal.}
     \label{fig8}
\end{figure*}

The GO enrichment of the integrated DMR set between the PCa tissues and the adjacent normal tissues is illustrated, as shown in Figure \ref{fig8}. It can be figured out that the DMGs obtained by DMRIntTk are enriched in pathways related to cell fate and pattern specification, skeletal system, and axonogenesis.

As we know, epigenetic reprogramming can lead to aberrant lineage specification and transition of tumor cells, which is closely associated with tumor initiation and progression. A study\cite{Li2020} showed that overexpression of ERG drives prostate cell fate reprogramming through the orchestration of chromatin interactions, which facilitates the function of ETS transcription factor (ERG) to promote luminal lineage differentiation. Luminal lineage differentiation further leads to luminal cell expansion, which is a structural feature of most PCa compared with normal prostate tissues.  Axonogenesis is a biological phenomenon that is crucial in the biology of PCa. Adriana et al.\cite{Adriana2014} corroborates that axonogenesis is involved in the biological process of the proliferation of PCa through activation of survival pathways and interaction with hormonal regulation. PCa is capable of metastasizing to osteoblasts and inducing extensive new bone deposition. In fact, bone is the most common site of metastasis for advanced solid tumors including PCa\cite{Janet2005}. It have been found that approximately $70\%$ of patients with advanced PCa will develop skeletal metastases\cite{Coleman1997}.

\subsubsection{GO enrichment analysis of the integrated DMR sets between the AD patients and normal controls}\label{subsec362}

\begin{figure*}
    \centering
    \includegraphics[scale=0.4]{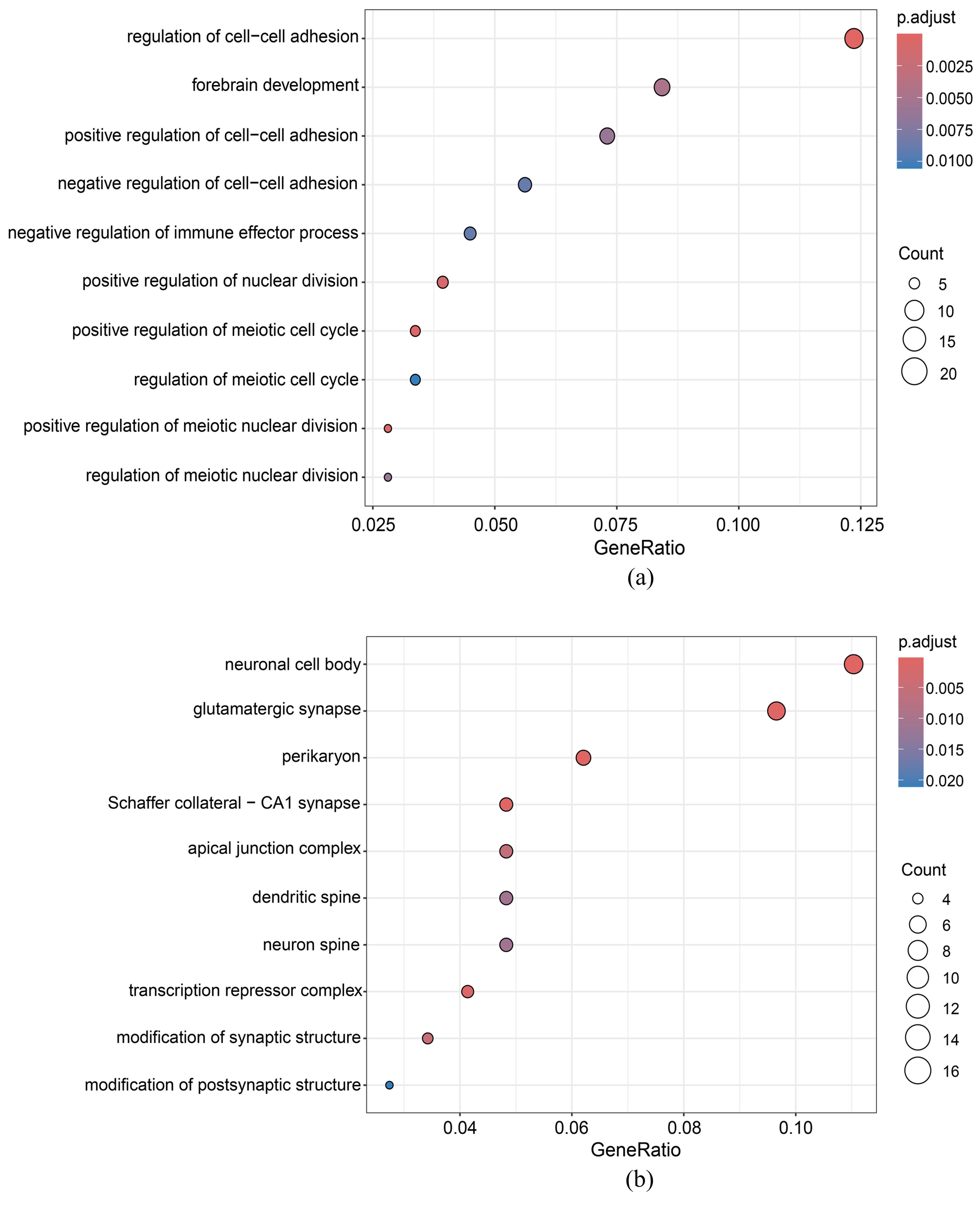}
    \caption{GO enrichment analysis of the integrated DMR set on AD versus normal in the (a) entorhinal cortex (EC), and (b) superior temporal gyrus (STG).}
    \label{fig9}
\end{figure*}

The  GO enrichment of the integrated DMR sets predicted between AD patientis and normal controls in the EC brain region is illustrated, as shown in Figure \ref{fig9}(a). It can be observed that the AD-associated DMGs in the EC are mainly enriched in pathways related to cell adhesion, cell nuclear division, and the meiotic cell cycle. Cell adhesion molecules play important roles in the core pathways of AD pathogenesis and progression, including A$\beta$ metabolism, cellular plasticity, neuroinflammation, and vascular changes. For example, expression of neural cell adhesion molecules (NCAM) is considered to be an indicator of neurogenesis, neuronal remodeling and plasticity. In a small study of patients with Parkinson's disease and AD, Guo et al.\cite{Guo2009} observed increased levels of soluble NCAM-120 splice variants in the CSF of AD patients compared to controls. Similarly, another study showed a trend towards increased levels of soluble NCAM in the CSF of AD patients compared to controls\cite{Helen2004}. The AD-associated DMGs in the EC are  also enriched in cell cycle-related pathways.  Exposure of the AD brain to extensive stress stimuli may trigger neuronal cell cycle termination. Many studies have shown that cell cycle proteins such as cyclins and CDKs are aberrantly expressed in AD brains\cite{Vincent1997,Smith1999,Busser1998}, and CDK inhibitors p16INK4a, p15INK4b, p18INK4c and p19INK4d have also been found to be aberrantly expressed in brain neurons of AD patients\cite{McShea1997,Arendt1996,Arendt1998}.

The GO enrichment of the integrated DMR sets predicted between AD patientis and normal controls in the STG brain region is illustrated, as shown in Figure \ref{fig9}(b). It can be found out that the AD-associated DMGs between AD patientis and normal controls in the STG  are mainly enriched in synapse-related pathways, including glutamatergic synapse, Schaffer collateral-CA1 synapse, modification of synpatic structure and modification of postsynpatic structure. Synapse loss, as a morphological feature that appear early in the onset of AD, has been found to be closely associated with the development of cognitive dysfunction in AD patients. Since calmodulin, particularly neuronal calmodulin (N-cadherins), is important for synapse formation and stability\cite{Nikolaos2009}, their possible role in AD pathology and clinical disease manifestations has become a hot research topic. Masliah et al.\cite{Masliah1993} oberserved that the loss of synaptic connections between neurons may facilitate the re-entry of cells into the cell division cycle, resulting in a perturbation of the cell cycle in AD patients. Thus, this available evidence suggests that cell cycle alterations play an crucial role in neurodegeneration in AD.

The AD-associated DMGs between AD patientis and normal controls in the STG  are also enriched in spine-related pathways, including dendritic spine and neuron spine. Dendritic spines (DS) are small, highly dynamic prominent structures on the dendritic membrane that form synapses\cite{Saravana2019}. These structures have specific sub-structural domains with specific functions in synaptic transmission and plasticity and DS are the main sites of structural modification of synaptic plasticity. In neurodegenerative diseases, dynamic morphological changes in the shape and density of dendritic spines can influence their functional features, leading to synaptic dysfunction and cognitive impairment. Peter et al.\cite{Peter2011} showed that dendritic spine dysfunction and subsequent synaptic failure are key features of the pathogenesis of AD.

\section{Discussion and Conclusions}\label{sec4}
By applying DMRIntTk to DMR sets predicted by seven methods in four scenarios, it demonstrates that DMRIntTk can be applied to different datasets with different methylation differences. The  methylation difference distributions shows that the integrated DMR sets have more large proportions of DMRs with higher methylation differences. Furthermore, the overlap analysis suggests that the integrated DMR sets include most DMRs predicted by all methods with methylation differences $\geq 0.5*m_{max}$.

In this paper, DMRIntTk was applied to integrate DMRs predicted from different methods on methylation array datasets.  DMRIntTk also can be applied to methylation profiles, extracted from bisulfite sequencing with some small modifications. Firstly, segment the genome with  a shorter distance between adjacent cytosines rather than 500bp,  since the density of cytosines in genome is much greater than that of  probes in methylation array. Further, in calculating the $DMRscore$,  the numbers of probes should be replaced by the number of cytosines.

There are several advantages of DMRIntTk in the integration of DMR sets from different methods. First, DMRIntTk can effectively trim the regions with small methylation differences in the original DMR sets by segmenting the genome into bins and weighting the bins based on the methylation differences and the reliability of covered methods. Secondly, DMRIntTk can be applied to different datasets with different methylation differences, due to  the automatic setting of two parameters, $c_t$ and $n_t$, in integrating the bins based on the DPC algorithm, which makes it possible to select the clustering thresholds according to the characteristics of a dataset. Therefore,  DMRIntTk can provide  a more comprehensive and reliable DMR set for downstream analysis.

\section{Declarations}
\paragraph{Ethics approval and consent to participate}
Not applicable.

\paragraph{Consent for publication}
Not applicable.
\paragraph{Availability of data and materials}
DMRIntTk has been implemented in a R package which is available at Github: \href{https://github.com/WjinZhang/DMRIntTk}{https://github.com/WjinZhang/DMRIntTk}.
The methylation array datasets involved in this  paper were downloaded from Gene Express Omnibus (GEO) with accession ids GSE48472, SE50192, GSE112047, GSE157272, GSE43414, GSE105109,  GSE125895, GSE43414, GSE66351, and GSE80970, and from the Religious Orders Study and Memory and Aging Project (ROSMAP) cohort[39] with Synapase ID syn3219045.
\paragraph{Competing interests}
The authors declare that they have no competing interests.

\paragraph{Funding}
This work was supported in part by the Natural Science Foundation of Hunan Province (No. 2022JJ30694 and No. 2022JJ30750); Central South University Innovation-Driven Research Programme (No. 2023CXQD065); Special Funds for Construction of Innovative Provinces in Hunan Province (NO. 2023GK1010).

\paragraph{Authors' contributions}
XP and YZ designed the density peak cluatering algorithm for integrating different DMR sets. XP, WZ, WC and WJ realized the algorithm. WZ preprocessed the methylation array datasets, applied the DMRIntTk on DMR sets predicted from four scenarios and evaluated the performance. WZ and XP were major contributors in writing the manuscript. GD, YZ and XP revised the manuscript. All authors read and approved the final manuscript.

\paragraph{Acknowledgements}
We are grateful for resources from the High Performance Computing Center of Central South University.

\nolinenumbers

%
%
%
\bibliographystyle{plos2015.bst}
\bibliography{references}

\end{document}